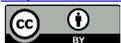

# Global Earthquake Prediction Systems

## Oleg Elshin[1], Andrew A. Tronin[2]

[1]President at Terra Seismic, Alicante, Spain/Baar, Switzerland
[2]Chief Scientist at Terra Seismic, Director at Saint-Petersburg Scientific-Research Centre for Ecological Safety of the Russian Academy of Sciences, St Petersburg, Russia
Email: oleg.elshin@terraseismic.com

## Abstract

Terra Seismic can predict most major earthquakes (M6.2 or greater) at least 2 - 5 months before they will strike. Global earthquake prediction is based on determinations of the stressed areas that will start to behave abnormally before major earthquakes. The size of the observed stressed areas roughly corresponds to estimates calculated from Dobrovolsky's formula. To identify abnormalities and make predictions, Terra Seismic applies various methodologies, including satellite remote sensing methods and data from ground-based instruments. We currently process terabytes of information daily, and use more than 80 different multiparameter prediction systems. Alerts are issued if the abnormalities are confirmed by at least five different systems. We observed that geophysical patterns of earthquake development and stress accumulation are generally the same for all key seismic regions. Thus, the same earthquake prediction methodologies and systems can be applied successfully worldwide. Our technology has been used to retrospectively test data gathered since 1970 and it successfully detected about 90 percent of all significant quakes over the last 50 years.

## 1. Introduction

Earthquake forecasting is one of the most ancient skills known to mankind. The first known forecast was made by Pherecydes of Syros about 2500 years ago: he made it as he scooped water from a well and noticed that usually very clean water had suddenly become silty and muddy. Indeed, an earthquake occurred two days later, making Pherecydes famous. Unusual natural phenomena, such as seismic clouds, changes in the behavior of living beings, etc., may be observed before a lot of major earthquakes. Ancient Greeks lived very close to nature and were able to detect unusual phenomena and forecast earthquakes. Multiple sources indicate that earthquake forecasting was a recognized science in ancient Greece.

From ancient Greece to the present day, countless scientists have tried to develop tools to predict earthquakes. Their attempts usually focused on searching for reliable precursors of forthcoming quakes. Nowadays, seismic and remote-sensing methods are considered to have the greatest potential in terms of resolving the earthquake prediction problem. Both of these leading methods can provide the real-time information necessary for global prediction. Scientists have already accumulated enough reliable historic data in order to perform accurate retrospective testing.

The most natural means of predicting earthquakes is by studying seismicity. Thus, seismic methods represent a first-choice group of methodologies for earthquake prediction. Various attempts have been made to predict earthquakes based on local seismicity analysis. One of the most outstanding achievements of the 1980s was the development of the family of M8 algorithms by Vladimir Keilis-Borok *et al.* [1]-[6]. However, a critical analysis of the forecasts it predicts has revealed a low success rate.

The second group of methodologies makes use of satellite remote sensing [5] [7] [8] [9]. A wide range of remote sensing methods has been applied to earthquake analysis and prediction such as visible and infra-red observations, satellite radar interferometry, and thermal surveying, among others. GPS methods have also been used. Data have been collated from several geophysical parameters to obtain a possible prediction signal: the Earth's surface, sea surface and air temperatures, outgoing longwave radiation (OLR) [10], humidity, surface displacement, etc. Numerous anomalies were recorded by many researchers before different earthquakes using remote sensing satellite techniques on land and sea surfaces and in the atmosphere.

## 2. Geophysics and Global Earthquake Prediction

The Terra Seismic global prediction methodology adheres to some widely accepted assumptions and is based on the observation and detection of real geophysical processes that always take place below the earth's surface before earthquakes [11]-[21]. While earthquakes happen very suddenly for humans, they are not sudden for nature. In nature, earthquakes build up over time in a gradual process involving the accumulation of a huge amount of physical stress. This accumulated stress is subsequently released as an enormous amount of energy when the earthquake strikes. For example, a magnitude 7 earthquake releases the same amount of energy as 32 Hiroshima bombs. Our method is based on the premise that such a huge accumulation of stress can be successfully detected well in advance. The area of future earthquake will be stressed, and, as a result, it will behave differently compared to other areas in the vicinity.

Actually, real earthquake prediction is very similar to the diagnosis of underlying human illnesses based on observing and analyzing each patient's signs and symptoms. Just like a fever is an indication to visit a doctor because something is wrong, anomalies in geophysical parameters indicate something is wrong in that specific part of the world. Before an earthquake strikes, we can notice anomalies in the data we collect. These deviations tell us that an event is coming. Major earthquakes are very rare but recurring events; they may take place in a specific area once every 30, 40, 50, 70 or even 100 years. As can be expected, we can observe historically unique combinations of certain parameters in stressed areas before they are hit by a forthcoming earthquake.

Major earthquakes require more time to develop and, therefore, can be detected earlier. For example, based on our current systems, we would have detected the following famous earthquakes with the lead times (detection time before the earthquake) given in Table 1.

In many cases, the greater the magnitude of the forthcoming event, the larger the area of observed stressed. The size of observed stressed areas roughly corresponds to estimates derived from the famous Dobrovolsky formula [22]. Also, fully in line with our expectations, we discovered that the size of the stressed area gradually increases during the longer buildup period associated with large magnitude events.

## 3. New Approach Based on Worldwide Big Data Acquisition and Processing

To develop our prediction method, our R&D team accumulated all the available data on earthquakes, their precursors and earthquake prediction from all sources over the last 2000 years. While some scientists concentrate most of their efforts and research on just one specific country or region, such as Japan or California, thanks to Big Data, our team can efficiently analyze data from the whole world in almost real time. We've discovered that reliable data for some specific regions, such as California, can be successfully complemented with data from other key seismic regions. So, by applying a global approach based on Big Data applications, we've managed to significantly increase the volume of reliable data for developing and testing completely novel theories, models and systems.

Table 1. Lead time periods: earthquake became detectable based on current prediction systems.

| No. | Date | Earthquake | Lead Time |
|-----|------|------------|-----------|
| 1 | 12-11-2017 | M7.3 Iran-Iraq | 4 months |
| 2 | 24-08-2014 | M6 Napa, California | 5 months |
| 3 | 07-09-2017 | M8.2 Chiapas, Mexico | 5 months |
| 4 | 15-10-2013 | M7.1 Bohol, Philippines | 10 months |
| 5 | 17-10-1989 | M6.9 Loma Prieta, California | 10 months |
| 6 | 06-04-2009 | M6.3 L'Aquila, Italy | 12 months |
| 7 | 14-11-2016 | M7.8 Kaikoura, New Zealand | 15 months |
| 8 | 05-07-2019 | M7.1 Ridgecrest, California | 2 years |
| 9 | 11-03-2011 | M9.1 Tohoku, Japan | 3 years |

We apply innovative satellite Big Data technology capable of processing and analyzing terabytes of information every day. The core of our system is mainly programmed in Python and lives in an open source software ecosystem. The frontend runs on the Apache web server and in the Debian GNU/Linux environment. The system core generally relies on NumPy for computations and a stack of open source geospatial tools. It uses GDAL to read and process archived satellite data, PROJ to compute the necessary coordinate transformations, and Matplotlib and CartoPy for the final graphics. Some parts of the systems were written in R language which is specifically targeted for statistical computations.

We selected the Python programming language so we could rapidly deploy the system in production and quickly implement new ideas and algorithms, thus constantly expanding the number of parameters involved. The open source Linux based software stack also affords us the ability to perform quick and painless deployment and adds greater flexibility when novel features are permanently added to the system.

Original software algorithms analyze satellite images and data from ground-based instruments in order to discover stressed areas based on area-specific comparisons of current data against historical data.

## 4. Terra Seismic Global Earthquake Prediction Technology Overview

Terra Seismic was established in 2012. In 2019, the company relocated to Baar, Switzerland. Terra Seismic can predict most major earthquakes (M6.2 or greater) at least 2 - 5 months before they will strike. We currently provide earthquake prediction for 25 key earthquake-prone regions: Alaska, the Balkans, California, Canada, the Caribbean, Central America, Chile, China, Greece, India and Pakistan, Indonesia, Iran, Italy, the Izu Islands, Japan, Kamchatka and the Kuril Islands, Mexico, Central Asia, New Zealand, Okinawa, Papua New Guinea, Peru, Philippines, Taiwan, and Turkey.

Terra Seismic can identify the epicenter of a forthcoming earthquake with a high level of confidence. At present, the company can identify potentially dangerous earthquake areas with a radius of 150 - 250 km. Generally, Terra Seismic does not promise to predict a quake if the earthquake's epicenter is located beyond a depth of 40 km. Some exceptions are possible. Fortunately, such quakes are almost always harmless, since the quake's energy is dissipated before reaching the Earth's surface. Sometimes, instead of a single forecasted earthquake, a few smaller earthquakes will strike in the specified area and time period. In such cases, Terra Seismic's systems correctly predict the sum total of the energy released by all the smaller seismic strikes. For this reason, the actual quake magnitude is sometimes less than the predicted one.

We develop long-term (from 2 to 5 years), mid-term (from 2 months to 2 years), and short-term (from 10 to 60 days) global prediction systems. The mid-term systems are the most reliable and can predict most major earthquakes at least 2 - 5 months in advance.

Terra Seismic technology has been used to retrospectively test data gathered since 1970, and its systems successfully predicted about 90 percent of all significant quakes over the last 50 years. Our technology has been in practical use since 2013. A top California earthquake insurer has been successfully testing our predictions for nearly five years after they became a Terra Seismic client in 2015. Terra Seismic is actively working on improving its prediction capabilities.

## 5. Remote Sensing

All kinds of thermal anomalies have repeatedly been registered before different earthquakes by many researchers using remote sensing satellite techniques applied to land and sea surface temperatures [7]. Thermal anomalies associated with strong earthquakes have been observed at various levels, from the ground surface up to the top of clouds. At present, the most promising is the Outgoing Longwave Radiation (OLR) anomaly measured at the top of clouds [10]. The advantage of this method is that it measures all of the infrared radiation emitted from the Earth's surface and atmosphere within the transparency window of 8 - 12 microns. OLR is currently mapped by the AIRS (Atmospheric Infrared Sounder) instrument launched into orbit in 2002. AIRS is one of six instruments on board the Aqua satellite, part of NASA's Earth Observing System.

## 6. Global Earthquake Prediction. Practical Cases

Let's demonstrate a few cases of real stress gradually accumulating before major earthquakes (Figure 1 and Figure 2). At some point the stressed area becomes detectable for our prediction systems. In many cases, the epicenter of a forthcoming earthquake is located near the center of the stressed area. However, in some cases, the epicenter is closer to the boundaries of the stressed area. A possible explanation is that the rupture zone represents a better indicator for major earthquakes (rather than the epicenter). The rupture may reach a length of 1300 km for M9 events [23].

## 7. Results

A few recent successful prediction cases are shown in Figures 3-6.

## 8. Conclusions

Terra Seismic can predict most major earthquakes (M6.2 or greater) at least 2 - 5 months before they will strike. Global earthquake prediction is based on determinations of the stressed areas that will start to behave abnormally before major earthquakes. The size of the observed stressed areas roughly corresponds to estimates calculated from Dobrovolsky's formula. To identify abnormalities and

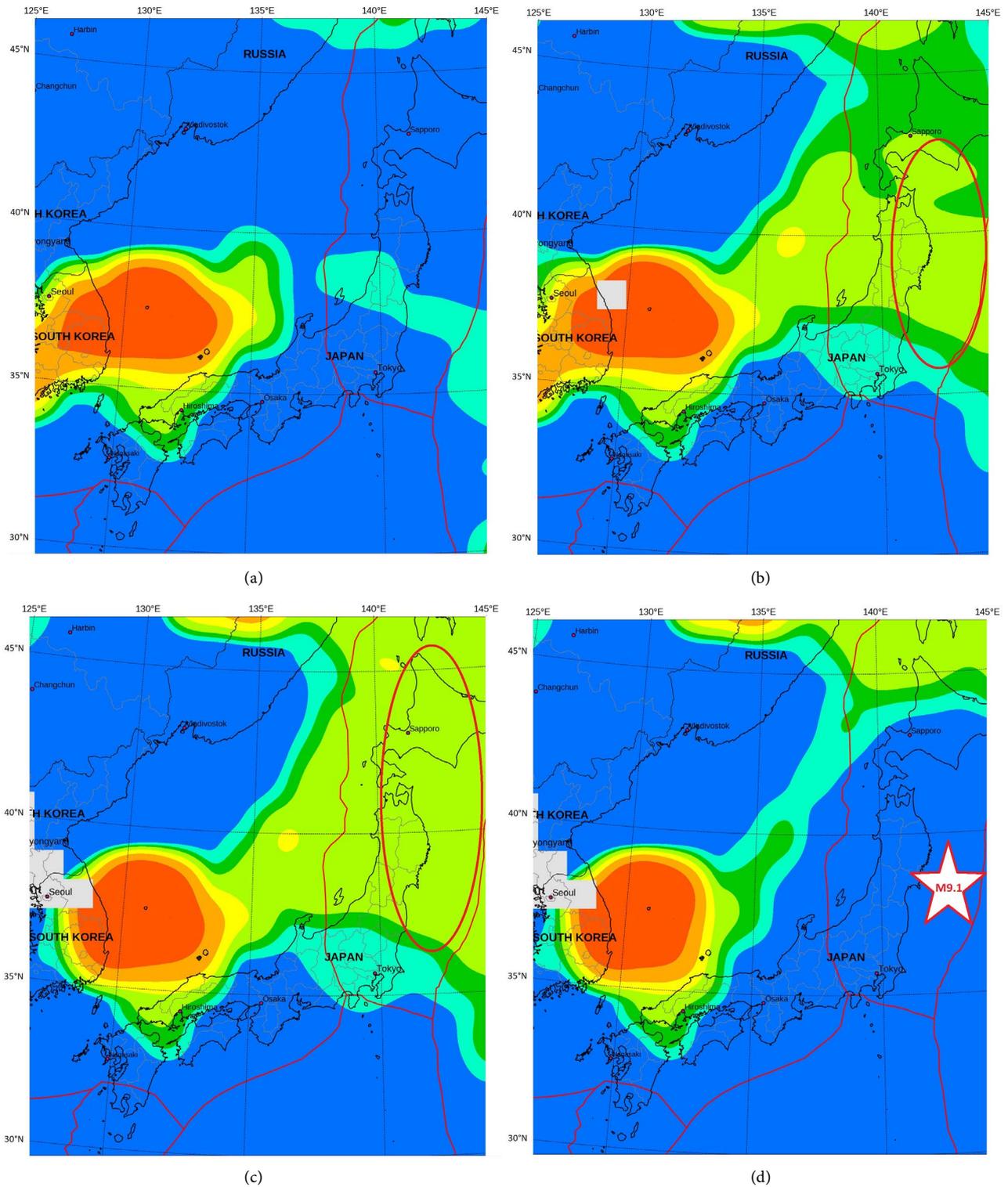

**Figure 1.** Example of prognostic signal analysis for M9.1 Tohoku earthquake 11 Mar 2011: (a) Mar 2010, (b) Aug 2010, (c) Mar 2011, (d) Apr 2011, after the shock. Note that the green stressed area grew larger in March 2011 due to an increased accumulation of stress. The red ellipse indicates the prognostic signal.

make predictions, Terra Seismic applies various methodologies, including satellite remote sensing methods and data from ground-based instruments. We currently

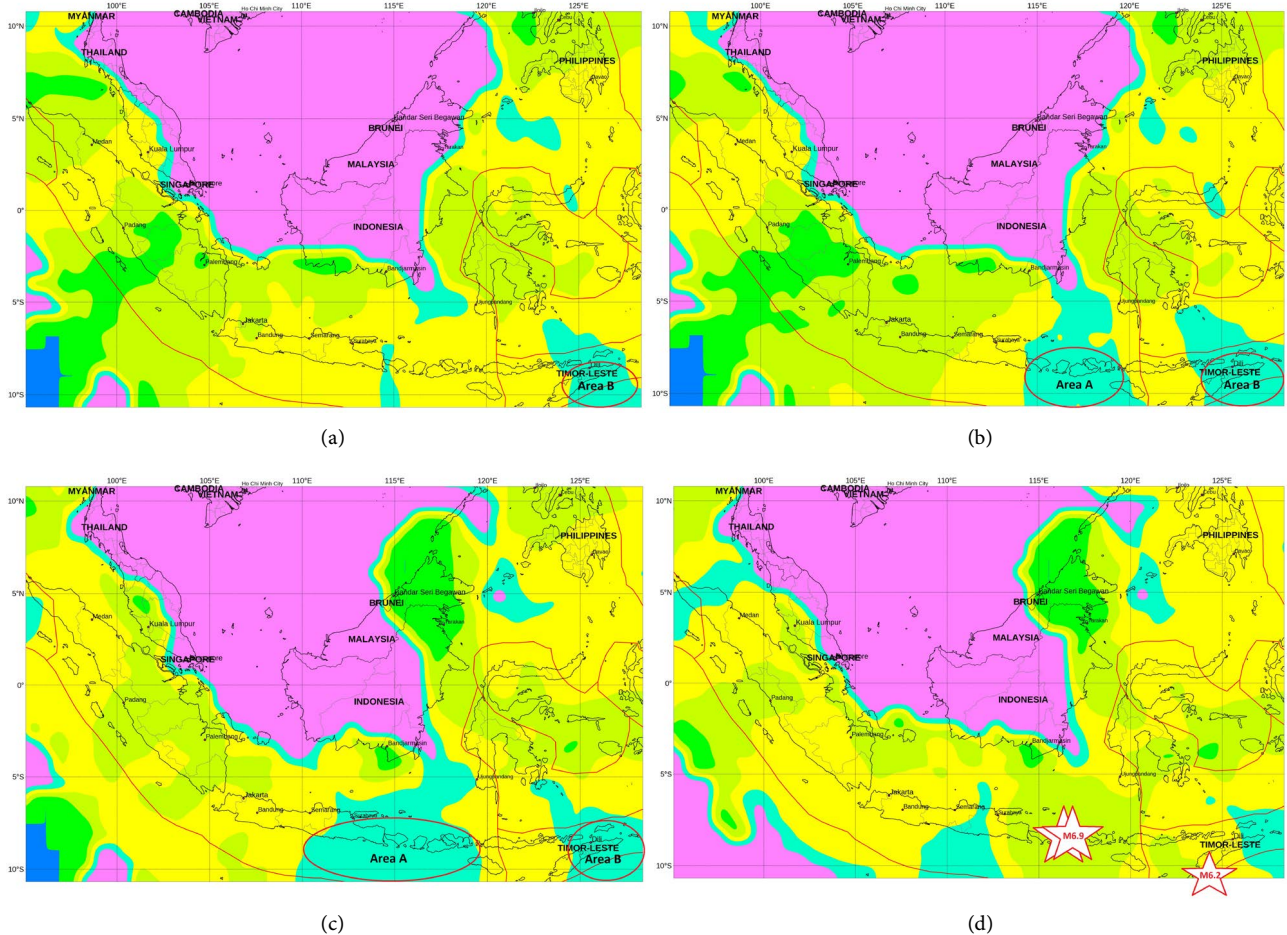

**Figure 2.** An example of how multiple earthquakes developed simultaneously in Indonesia in 2018. Area A: preparation of M6.4 quake on 28.07.2018, M6.9 quake on 05.08.2018, M6.3 quake and M6.9 quake on 19.08.2019 in Lombok Region. Area B: preparation of M6.2 quake on 28.08.2018 in the East Timor region. (a) Dec 2017, (b) Jan 2018, (c) Jul 2018, (d) Sep 2018, after the shocks. Note that the cyan stressed areas grew larger due to an increased accumulation of stress. The red ellipse indicates the prognostic signal.

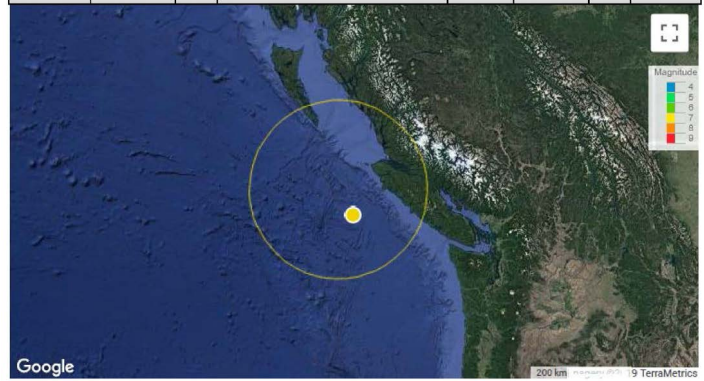

**Figure 3.** Example of prediction and real quake comparison for 22.10.2018 M6.8 earthquake in Vancouver Island, Canada region. Yellow circle indicates prognostic area and yellow dot shows the location of epicenter.

Terra Seismic : Predicted earthquake in Banda Sea

| | Date | Time, UTC | Region, Country | Latitude | Longitude | M | Waiting period, days |
|---|---|---|---|---|---|---|---|
| Alert | 14.10.2018 | 11:07 | Kepulauan Barat Daya, Indonesia/Banda Sea | 7.4S±3.0 | 128.2E±3.0 | 7 | 1-365 days |
| Real Quake | 24.06.2019 | 2:53 | Banda Sea | 6.4S | 129.2E | 7.3 | 252 days |

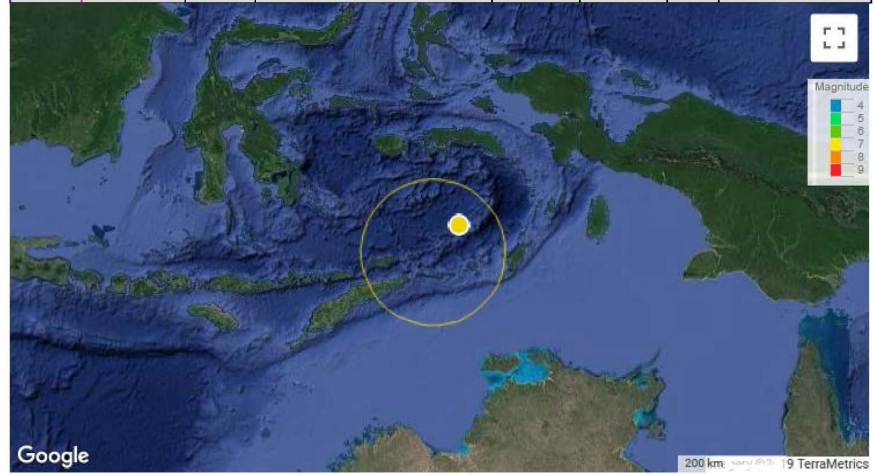

**Figure 4.** Example of predicted M7.3 earthquake in Banda Sea, Indonesia. Yellow circle indicates prognostic area and yellow dot shows the location of epicenter.

Terra Seismic : Predicted earthquake in Southwest of Sumatra

| | Date | Time, UTC | Region, Country | Latitude | Longitude | M | Waiting period, days |
|---|---|---|---|---|---|---|---|
| Alert | 17.07.2019 | 12:36 | Southern Sumatra/Sunda Strait, Indonesia | 6.25±2.5 | 104.7E±2.5 | 7+ | 1-365 days |
| Real Quake | 02.08.2019 | 12:03 | Southwest of Sumatra, Indonesia | 7.3S | 104.8E | 6.9 | 16 days |

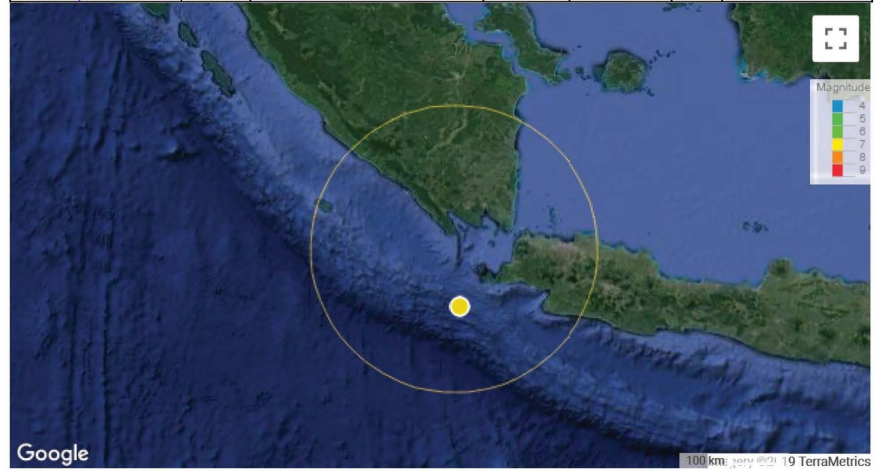

**Figure 5.** Example of predicted M6.9 earthquake in Southwest of Sumatra, Indonesia. Yellow circle indicates prognostic area and yellow dot shows the location of epicenter.

process terabytes of information daily, and use more than 80 different multiparameter prediction systems. Alerts are issued if the abnormalities are confirmed by at least five different systems. We observed that geophysical patterns of earthquake development and stress accumulation are generally the same for all key seismic regions. Thus, the same earthquake prediction methodologies and systems can be applied successfully worldwide.

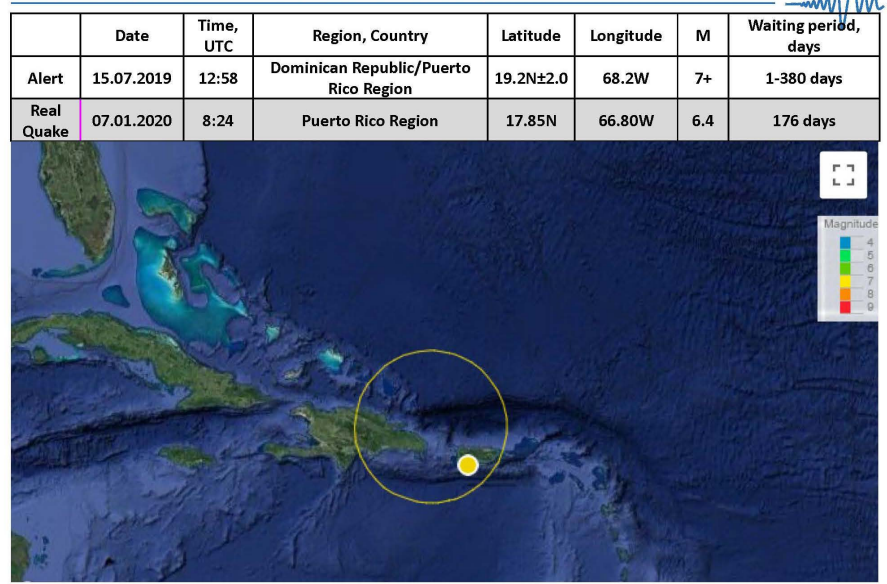

**Terra Seismic : Predicted earthquake in Puerto Rico**

| | Date | Time, UTC | Region, Country | Latitude | Longitude | M | Waiting period, days |
|---|---|---|---|---|---|---|---|
| Alert | 15.07.2019 | 12:58 | Dominican Republic/Puerto Rico Region | 19.2N±2.0 | 68.2W | 7+ | 1-380 days |
| Real Quake | 07.01.2020 | 8:24 | Puerto Rico Region | 17.85N | 66.80W | 6.4 | 176 days |

**Figure 6.** Example of predicted M6.4 earthquake in Puerto Rico region. Yellow circle indicates prognostic area and yellow dot shows the location of epicenter.

Stress gradually accumulates before a major earthquake. To measure the different stages of stress accumulation, we have developed long-term (from 2 to 5 years), mid-term (from 2 months to 2 years), and short-term (from 10 to 60 days) global prediction systems. The most reliable are the mid-term systems that can predict most major earthquakes at least 2 - 5 months in advance. In some cases we can determine the final stage of stress build-up. We can also predict the epicenter of a forthcoming earthquake with a high degree of confidence to within a radius of 150 - 250 km. Terra Seismic currently provides earthquake predictions for 25 key earthquake-prone regions. Our technology has been used to retrospectively test data gathered since 1970 and it successfully detected about 90 percent of all significant quakes over the last 50 years. Throughout 2017-2020, Terra Seismic's work was presented to more than 150 university professors from 63 countries. Our technology has been in practical use since 2013.

Our paramount priority is to help governments save human lives. Terra Seismic calls for collaboration with all governments and agencies responsible for dealing with natural disasters.

## Acknowledgements

This project was not possible without the scientific data provided by different government agencies, international organizations, science institutions and academia. We would like to acknowledge their leading contribution to Earth and space data collection.

We wish thank to US Geological Survey (USGS), European-Mediterranean Seismological Centre (EMSC), Japanese Meteorological Agency (JMA), National Aeronautical and Space Administration (NASA), National Oceanic and Atmos-pheric Administration (NOAA), European Space Agency (ESA), International GNSS Service (IGS), Jet Propulsion Laboratory (JPL)/Caltech, Ionospheric Prediction Service (IPS), Weather Underground and World Data Center (WDC) in Kyoto, Japan.

## Conflicts of Interest

The authors declare no conflicts of interest regarding the publication of this paper.